\documentclass[twocolumn,draft]{svjour3}            
\usepackage{amsmath,amsfonts,amssymb,amscd,graphicx}
\usepackage{cancel}

\newenvironment{Proof}{\par\noindent{\bf Proof.}}{\hfill $\Box$\par\smallskip}
\def\demo#1{\begin{Proof}}

\def\enddemo{\end{Proof}}

\font\Frak=eusm10

\def\vett#1{\underline{#1}}
\def\e_#1{\vett e_#1}
\def\k_#1{\vett k_#1}
    \def\cost{\text{cost.}}

\def\a{\alpha}
\def\b{\beta}

\def\l{\lambda}

\def\s{\sigma}
\def\th{\vartheta}

\def\vphi{\varphi}
\def\w{\omega}
\def\x{\hat{\plus50 x\@}}
\def\z{\zeta}
\def\xtilde{\raise-1ex\hbox to0pt{$\scriptstyle\sim\hss$} \@\?x{}}
\def\hxtilde{\raise-1ex\hbox to0pt{$\scriptstyle\sim\hss$} \@\x{}}
\def\A{\mathcal A}
\def\B{\mathfrak B}

\def\E{\mathfrak E}
\def\Eps{\text{\Frak E}}
\def\F{\mathfrak F}
\def\H{\mathcal H}

\def\Routh{\mathcal R}
\def\Surf{\mathcal S}
\def\W{\Omega}

\def\Frac#1#2{\text{\large$\frac{#1}{#2}$}}
\def\de#1/de#2{\frac{\partial{#1}}{\partial{#2}}}
\def\d#1/d#2{\frac{d#1}{d#2}}
\def\sd#1/de#2/de#3{\ifx#2 \frac{\plus02\partial^{\@\@2}#1}{\plus90\partial\@#3^{\@2}}%
\else\frac{\plus02\partial^{\@\@2}#1}{\partial\?#2\@\partial\?#3}\fi}
\def\oD#1/d#2{\textstyle{\text{\large$\d{#1}/d{#2}$}}}
\def\De#1/de#2{\textstyle{\text{\large$\de{#1}/de{#2}$}}}
\def\SD#1/de#2/de#3{\ifx#2 \frac{\plus02\partial^{\@\@2}#1}
    {\plus90\partial\@#3^{\@2}} \else\frac{\plus02\partial^{\@\@2}#1}
    {\partial\@#2\@\partial\@#3}\fi}
\def\Sd#1/de#2/de#3{\textstyle{\text{\large$\sd{#1}/de{#2}/de{#3}$}}}
\def\SDs#1/de#2/de#3{\frac{\plus02\partial^{\@2}#1}
    {\partial\@#2\@\partial\@#3}}
\def\DE{\partial}

\def\TE{T\!\@S\!\@\?E\/(3)}

\def\tondo{$\phantom{.}\bullet\;\;$}

\def\And{,\@\ldots\hskip-.4pt,}

\def\dv{d^{\@\?3}\vett x}
\def\AWP{A_\text{\@\tiny WP}}
\def\GMT{G\!\@M_\text{\@\tiny T}}
\def\GML{G\!\@M_\text{\@\tiny L}}
\def\FB{\vett F^{\text{\@\tiny(B)}}}
\def\MB_#1{\vett M^{\text{\@\tiny(B)}}_{#1}}
\def\QB_#1{Q^{\text{\@\tiny(B)}}_{#1}}
\def\UB{U^{\text{\?\tiny(B)}}}
\def\UG{U^{\text{\?\tiny(weight)}}} \def\QG_#1{Q_{\,#1}^{\text{\?\tiny(weight)}}}
\def\piB{\pi^{\text{\?\tiny(B)}}}
\def\xC{x_{\text{\?\tiny C}}}   \def\yC{y_{\text{\?\tiny C}}}   

\def\grad{\operatorname{grad}}

\def\div{\operatorname{div}}
\def\vol{\operatorname{vol}}

\def\@{\hskip.65pt}
\def\?{\hskip.3pt}
\def\plus#1#2{\vrule height#1pt width0pt depth#2pt}

\begin{document}

\title{Floating rigid bodies: a note on the conservativeness of the hydrostatic effects}

\author{Enrico Massa \and Stefano Vignolo}

\institute{
            E. Massa \at DIME, Sez.~Metodi e Modelli Matematici, Universit\`a di Ge\-no\-va. Piazzale Kennedy, Pad.~D. 16129 Genova (Italy). \\\email{massa@dima.unige.it}
    \and
          S. Vignolo \at DIME, Sez.~Metodi e Modelli Matematici, Universit\`a di Ge\-no\-va. Piazzale Kennedy, Pad.~D. 16129 Genova (Italy). \\\email{vignolo@dime.unige.it}
}

\date{Received: date / Accepted: date}

\maketitle
\maketitle
\abstract%
{Within the framework of Lagrangian mechanics, the conservativeness of the hydrostatic forces acting on a floating rigid body is proved. The representation of the
associated hydrostatic potential is explicitly worked out. The invariance of the resulting Lagrangian with respect surge, sway and yaw motions is used in connection with
the Routh procedure in order to convert the original dynamical problem into a reduced one, in three independent variables. This allows to put on rational grounds the
study of hydrostatic equilibrium, introducing the concept of {\it pseudo--stability}, meant as stability with respect to the reduced problem. The small oscillation of
the system around a pseudo--stable equilibrium configuration are discussed.}

%
\PACS{47.85.Dh \and 45.20D \and 45.20.Jj}

\date{}

\section{Introduction}\label{S1}
Hydrostatics plays a central role in many fields of applied sciences and engineering. The first contributions go back to the celebrated studies of Archimedes, who first
formulated the basic laws of the discipline.

In spite of this long history, some theoretical aspects still lack a rigorous mathematical setting.

This  happens e.g.~in the so called \emph{metacentric stability analysis\/}, where the evolutions originated by small displacements of a vessel from a state of
hydrostatic equilibrium are assumed to remain ``small'', thus justifying the linearization of the exact dynamical equations.
The linearized equations are then used in order to discuss the stability of the original configuration.

The resulting conclusions are unquestionably supported by practical and experimental evidence. From a theoretical viewpoint, however, a rigorous approach should rather
follow the inverse logical path: the stability of the equilibrium configuration should be analysed first, in order to motivate the replacement of the exact equations
with the linearized ones.

A powerful tool in this sense would be the validity of a conservativeness theorem, indicating that the generalized forces associated with buoyancy are derivable from a
suitable \emph{potential\/}.
This is obviously true for rigid bodies totally immersed in a fluid, the hydrostatic effects being in that case equivalent to a constant \emph{buoyant force\/}, applied
to a body--fixed \emph{buoyancy center\/}.

In the case of floating bodies, matters are complicated by the fact that the submerged volume, and therefore also the associated dynamical effects, depend the
configuration of the system.

This brief note is devoted to an analysis of this point. The central result is the conservativeness of the hydrostatic forces on a floating body, as well as the
representation of the corresponding potential.

The analysis yields back the standard results concerning the absence of restoring forces in the directions of surge, sway and yaw motions, relating them to the
invariance properties of the Lagrangian of a rigid body floating in calm water.

The consequences of this invariance are further elaborated, making use of a classical algorithm, known as the \emph{Routh procedure\/} \cite{Goldstein}: through the
latter, the original dynamical problem is reduced to a simpler one, involving only the significant (non--cyclic) variables.

This helps introducing a new concept, here called \emph{pseudo--stability\/}, which proves to be the natural one in connection with the study of the equilibrium
configurations of a floating rigid body.

The small oscillations of a vessel around a pseudo--stable equilibrium configuration are finally discussed. Although elementary, the results may have some interest in
practical applications.

\section{Hydrostatic potential}\label{S2}
Let $\@\B\@$ denote a floating rigid body (``the ship''), moving under the action of weight and buoyancy forces. For simplicity, $\@\B\@$ is assumed to possess a
longitudinal symmetry plane $\@\Pi\@$, as it happens in the case of standard vessels, presenting the usual port--starboard symmetry.

We denote by $D$ the submerged part of $\B$, by
\linebreak
$V=\vol(D)$ the submerged volume, by $G$ the mass center of $\B$, by $\@B\@$ the buoyancy center, by $\@\FB$ the resultant of the hydrostatic forces, and by $\MB_G$ the
associated torque, relative to $G$.

All quantities, when referred to the equilibrium configuration, will be marked by an asterisk: in this way, the archimedean condition reads $\@m=\rho\/V^*$, with $\@m\@$
and $\@\rho\@$ respectively denoting the mass of $\@\B\@$ and the mass density  of the fluid.

The intersection of $\@\B\@$ with the horizontal plane $\@\Sigma\@$ representing the free surface of the fluid, henceforth denoted by $\@A\@$, is called the
\emph{waterplane\/}; its equilibrium counterpart $\@A^*\@$, viewed as a body--fixed object, is called the \emph{static waterplane\/}.


For descriptive purposes, we adopt a fixed cartesian frame $\@\F=\big\{\W,\k_1,\k_2,\k_3\big\}\@$\vspace{1pt}, with coordinates $\@\xi,\eta,\z\@$, coordinate plane
$\@\z=0\@$ coinciding with $\@\Sigma\@$ and $\@\k_3\@$ axis pointing downward.
We also consider a body--fixed cartesian frame $\@\F'=\big\{G,\e_1,\e_2,\e_3\big\}\@$, with origin at the mass center $\@G\@$ and coordinates $\@x_1,x_2,x_3\@$ chosen in
such a way as to make the plane $\@x_2=0\@$ identical to the longitudinal symmetry plane $\@\Pi\@$, the plane $\@x_3=0\@$ parallel to the static waterplane $\@A^*\@$ and
the $\@\e_3\@$ axis pointing in the same half--space as $\@\k_3\@$.

The orthogonal projection of $\@G\@$ on $\@A^*\@$ is indicated by $\@\bar G\@$; the distance $\@|(\bar G-G)|\@$ is denoted by $\@d\@$. The time derivatives of vectors in
the frames $\@\F\@,\@\F'\@$ are respectively denoted by $\@\Big(\oD/dt\Big){\plus05\!}_\F\@$ and $\@\Big(\oD/dt\Big){\plus05\!}_{\F'}\@$.

The configurations of $\@\B\@$ are parameterized by the coordinates $\@\xi,\eta,\z\@$ of $\@G\@$ and by the Bryan Tait angles $\@\psi,\th,\vphi\@$ (yaw, pitch and roll).
The relation between the bases $\@\k_i\@$ and $\@\e_i\@$ is summarized into the orthogonal matrix $\@R_{ij}:=\k_i\cdot\e_j\@$. In particular, for later use, we recall
the expression \cite{Fossen,Lewandowski}
\begin{equation}\label{2.1}
\k_3=-\sin\th\,\e_1+\cos\th\@\sin\vphi\,\e_2+\cos\th\@\cos\vphi\,\e_3
\end{equation}

Letting $\@I_{ij}=I_G\/(\e_i)\cdot\e_j\@$ and $\@\w_i:=\vett\w\cdot\e_i\@$\vspace{1pt} respectively denote the components of the inertia tensor and of the angular
velocity of $\@\B\@$ in the body--fixed basis, the evolution of $\@\B\@$ is determined by the Lagrange equations
\begin{equation}\label{2.2}
\d/dt\,\de L'/de{\dot q^k}\,-\,\de L'/de{q^k}\,=\,\QB_k 
\end{equation}
with the ``partial Lagrangian''
\begin{multline}\label{2.3}
L'\@=\@T\@+\@\UG\@=\\=\@\tfrac12\,m\big(\dot\xi^2+\@\dot\eta^2+\@\dot\z^2\big)\,+\@\tfrac12\,I_{ij}\,\w_i\,\w_j\,+\,m\@g\,\z
\end{multline}
embodying the potential of the weight force, and the generalized forces $\@\QB_k\@$ expressing the buoyancy effects.\vspace{1pt}

To evaluate the latter, taking the positional character of the hydrostatic forces into account, we first consider the associated power $\@\piB$, and then apply the
relation $\@\QB_k=\@\De\piB/de{\dot q^k}\@$.
In detail, denoting by $\@p\@$ the hydrostatic pressure and recalling the identities
\begin{equation*}
\vett v_P=\vett v_G+\vett\w\wedge(P-G)\@,\quad \div\vett v_P=0\@,\;\quad \grad p=\@\rho\@g\@\k_3
\end{equation*}
we have
\footnote%
{Needless to say, the expression \eqref{2.4} is identical to $\@\piB=\FB\!\cdot\vett v_G+\MB_G\!\cdot\vett\w\@$\vspace{1pt}.}
\begin{multline}\label{2.4}
\piB\@=-\int_{\DE\!\@D}p\,\vett n\cdot\vett v_P\,d\?S\@=\@-\int_D\div\@(\?p\@\vett v_P)\@\dv\@=                                             \\[3pt]
=\@-\int_D\@\grad p\cdot\vett v_P\,\dv\@=\@-\@\rho\@g\int_D\@\k_3\cdot\vett v_P\,\dv
\end{multline}
On the other hand, the identifications $\,(P-G)=x_i\,\e_i\@$, $\@\vett v_G=\dot\xi\,\k_1+\dot\eta\,\k_2+\dot\z\,\k_3\@$, $\,R_{3i}=\k_3\cdot\e_i$ entail the identity
\begin{multline}\label{2.5}
\k_3\cdot\vett v_P=\dot\z\@+\@\k_3\cdot\vett\w\wedge\e_i\,x_i=\@\dot\z\@+\@\k_3\cdot\!\biggl(\d\e_i/dt\biggr){\plus07\!}_\F\@x_i=           \\[3pt]
=\@\dot\z\@+\@\d/dt\,\big(\k_3\cdot\e_i\big)\,x_i\@=\@\dot\z\@+\@\d\/R_{3i}/dt\;x_i\;
\end{multline}
whence also
\begin{equation}\label{2.6}
\piB\!=-\rho\@g\int_D\@\biggl(\dot\z\@+\@\d\/R_{3i}/dt\,x_i\biggr)\@\dv
\end{equation}
\begin{multline}\label{2.7}
\QB_k=-\rho\@g\int_D\@\biggl(\de\dot\z/de{\dot q^k}+\de/de{\dot q^k}\,\d\/R_{3i}/dt\;x_i\biggr)\@\dv=                                       \\[3pt]
=-\rho\@g\int_D\@\biggl(\de\z/de{q^k}\@+\@\de R_{3i}/de{q^k}\;x_i\biggr)\@\dv
\end{multline}

\begin{subequations}\label{2.8}
Adopting the notation $\@\QB_\xi,\QB_\eta\And\QB_\vphi$\vspace{1pt} in place of the anonymous one $\@\QB_1\And\QB_6$,\vspace{1pt} a straightforward comparison of
eq.~\eqref{2.7} with eq.~\eqref{2.1} yields the expressions
\begin{align}
& \QB_\xi=\QB_\eta=\QB_\psi=0\,,\quad\;\QB_\z=-\@\rho\@g\int_D\@\dv                                                                                 \label{2.8a}\\
& \QB_\th\@=\@\rho\@g\int_D\@\big[\cos\th\,x_1\@+\@\sin\th\@(\sin\vphi\,x_2\@+\nonumber\\[-5pt]&\hskip4.7cm+\@\cos\vphi\,x_3)\big]\@\dv               \label{2.8b}\\
& \QB_\vphi\@=\@-\@\rho\@g\int_D\@\cos\th\@(\cos\vphi\,x_2-\@\sin\vphi\,x_3)\@\dv                                                                   \label{2.8c}
\end{align}
\end{subequations}

A fairly more important result is expressed by the following
\begin{theorem}\label{Teo2.1}
The hydrostatic effect is conservative.
\end{theorem}
\begin{Proof}
To start with, we rephrase eq.~\eqref{2.6} in the equivalent form
\begin{multline}\label{2.9}
\piB\!=\rho\@g\biggl\{-\@\d/dt\@\int_D\big(\z+ R_{3i}\,x_i\big)\,\dv\;+                                                \\
+\z\;\d/dt\int_D\@\dv\,+\,R_{3i}\;\d/dt\int_D\@x_i\,\dv\,\biggr\}
\end{multline}

The plan is to evaluate the last two integrals in the body--fixed frame $\@\F\@'$, where the integrands are independent of the configuration variables $\@q^k$.

To this end we notice that, in the frame $\@\F\@'$, the boundary $\@\DE\!\@D\@$ consists of a part at rest $\@\overline{\DE\!\@D}$, identical to the submerged boundary
of $\@\B\@$, and of the waterplane $\@A\@$, variable in relation to the motion of $\@\B\@$.

More specifically, denoting by $\@\E_3\@$ the group of rigid motions in $\@E_3\@$\vspace{1pt} and by $\@\E_2\subset\E_3\@$ the subgroup of transformations preserving the
plane $\@\z=0\@$, it is readily seen that the elements of $\@\E_2\@$ do not modify the domain $\@D\@$, and that two elements $\@\s,\@\tau\in\E_3\@$ belonging to the same
coset of $\@\E_2\@$, i.e.~satisfying $\@\tau\cdot\s^{-1}\in\E_2\@$, affect $\@D\@$ in the same way.

From this, denoting by $\@\vett v_P=\vett v_G+\vett\w\wedge(P-G)\@$ the velocity of the points of $\@\B\@$ in the fixed frame of reference $\@\F\@$, and by $\@\vett X\@$
the speed of deformation of the boundary $\@\DE\!\@D\@$ in the frame $\@\F\@'$, we conclude that $\@\vett X\@$ vanishes on $\@\overline{\DE\!\@D}\@$ and is identical to
$\@-\@(\vett v_P\cdot\k_3)\,\k_3\@$ (namely to the opposite of the projection of $\@\vett v_P\@$ in the direction $\@\k_3\@$) at each point $\@P\in A\@$.\vspace{1pt}

By the transport equation, recalling eq.~\eqref{2.5}, as well as the fact that the outgoing unit normal on $\@\DE\!\@D\@$ coincides with $\@-\@\k_3\@$ along $\@A\@$, we
have then the evaluation
\begin{align*}
\z\;\d/dt\int_D&\@\dv\,+\,R_{3i}\;\d/dt\int_D\@x_i\,\dv\@=                                                                                                 \\[3pt]
&=\int_A\big(\z+ R_{3i}\,x_i\big)\@\vett v_P\cdot\k_3\,d\?S\@=                                                                                           \\[3pt]
&=\int_A\big(\z+ R_{3i}\,x_i\big)\biggl(\dot\z+ \d/dt\,R_{3i}\,x_i\biggr)\/d\?S\@=                                                                       \\[3pt]
&=\frac12\int_A\@\d/dt\@\big(\z+ R_{3i}\,x_i\big)^2\@d\?S
\end{align*}

Finally, given any function $\@f\/(q^k,x_i)\@$, we observe that the two--dimensional transport equation, applied to the integral $\@\int_A\@f\,d\?S\@$, entails the
relation
\begin{equation*}
\d/dt\@\int_A\@f\/(q^k\/(t),x_i)\,d\?S\,=\/\int_A\@\de f/de{q^k}\,\dot q^k\,d\?S\,+\/\int_{\DE\!\@A}\@f\,\cancel{\vett X\cdot\plus85 \vett n}\,\@d\?l
\end{equation*}
where, as above, $\@\vett X=-\@(\vett v_P\cdot\k_3)\,\k_3\@$ denotes the velocity of deformation of the boundary $\@\DE\!\@A\@$ in the frame $\@\F\@'$, while $\@\vett
n\@$, here representing the outgoing unit normal to the boundary $\@\DE\!\@A\@$, is a vector belonging to the plane $\@\z=0\@$, orthogonal to $\@\k_3\@$.

On account of the stated results, eq.~\eqref{2.9} may be written in the final form
\begin{multline*}
\piB\!=\rho\@g\,\d/dt\@\biggl\{-\int_D\@(\z+ R_{3i}\,x_i)\,\dv\,+\\+\frac12\int_A\big(\z+ R_{3i}\,x_i\big)^2\@d\?S\biggr\}
\end{multline*}
showing that the hydrostatic effect is indeed conservative, with potential
\begin{multline}\label{2.10}
\UB\@=\,\rho\@g\,\biggl\{-\int_D\@(\z+ R_{3i}\,x_i)\,\dv\,+\\+\frac12\int_A\big(\z+ R_{3i}\,x_i\big)^2\@d\?S\biggr\}
\end{multline}
\end{Proof}

\begin{remark}\label{Rem2.2.1}\rm
With the stated choices of the coordinates, we have the identification
\begin{multline}\label{2.11}
\z+ R_{3i}\,x_i\@=\@\z+ \k_3\cdot\e_i\,x_i\@=\@\k_3\cdot\big[\@(G-\W)+(P-G)\big]\@=\\=\@\k_3\cdot(P-\W)\@=\@\z\/(P)\;\,
\end{multline}

Since, as explicitly assumed, the origin $\@\W\@$ of the fixed frame of reference is placed on the free surface of the fluid, eq.~\eqref{2.11} implies $\,\z+
R_{3i}\,x_i\/(P)=0\@,\;\,\forall\@P\in A\@$.

Denoting by $\@B\@$ the buoyancy center, eq.~\eqref{2.10} reduces then to the simpler and intuitively more appealing expression
\begin{multline}\label{2.12}
\UB\@=\,-\rho\@g\@\int_D\@(\z+ R_{3i}\,x_i)\,\dv\,=\\=\,-\rho\@g\@\int_D\@\z\/(P)\,\dv \,=\; -\rho\@g\@V\@\z\/(B)
\end{multline}

In the case of a totally immersed body, eq.~(\ref{2.12}) is exactly what one would expect on elementary grounds.

The interesting fact is that, as long as $\@\W\@$ is chosen on the free surface of the fluid, the same expression \eqref{2.12} holds for an arbitrary floating body, with
$\@B\@$ and $\@V$ depending on the configuration variables.

Of course, it must be borne in mind that, unlike eq.~\eqref{2.10}, the representation (\ref{2.12}) is not invariant under vertical translations of the origin of the
fixed frame.
\end{remark}

\section{Routh procedure and pseudo-stable equilibrium configurations}\label{S3}
On account of Theorem \ref{Teo2.1}, the dynamical behaviour of a floating rigid body is completely described by a Lagrangian of the form
\begin{multline}\label{2.13}
L\,=\,T\@+\@\UG\@+\@\UB\@=\\=\tfrac12\@m\big(\dot\xi^2+\dot\eta^2+\dot\z^2\big)+\tfrac12\@I_{ij}\,\w_i\,\w_j+m\@g\,\z\@+\UB
\end{multline}
with $\@\UB$ given by eq.~\eqref{2.10}, and the angular velocity $\@\vett\w:=\w_i\@\e_i\@$ expressed in terms of the Bryan Tait angles by the equation (see
\cite{Fossen,Lewandowski} for details)
\begin{multline}\label{2.14}
\vett\w=\big(\dot\vphi -\dot\psi\@\sin\th\big)\@\e_1+\big(\dot\psi\@\cos\th\@\sin\vphi+\dot\th\@\cos\vphi\big)\,\e_2+\\+
\big(\dot\psi\@\cos\th\@\cos\vphi-\dot\th\@\sin\vphi\big)\@\e_3
\end{multline}

On account of eqs.~\eqref{2.10}, \eqref{2.14}, the variables $\@\xi,\eta,\psi\@$ are \emph{cyclic\/} in the Lagrangian \eqref{2.13}, thus ensuring the conservation of
the kinetic momenta
\begin{equation}\label{2.15}
p_\xi\@=\@m\@\dot\xi\,,\quad\, p_\eta\@=\@m\@\dot\eta\,,\quad\, p_\psi\@=I_{ij}\,\w_i\,\de\@\w_j/de{\plus{10}0\dot\psi}\qquad
\end{equation}

A standard reduction technique, known as \emph{the Routh procedure\/}, may then be applied in order to dig out of the Lagrange equations a subsystem of three
differential equations for the determination of the unknowns $\@\z\/(t),\vphi\/(t),\th\/(t)\@$.
The idea is well known \cite{Goldstein}: the conserved momenta are adopted in place of the jet--coordinates $\@\dot\xi,\dot\eta,\dot\psi\@$ as independent variables in
the velocity space, the soundness of the procedure being ensured by the solvability of eqs.~\eqref{2.15} with respect to $\@\dot\xi,\dot\eta,\dot\psi\@$.

Setting $\@q^\a=\{\z,\th,\vphi\}\@$, $\@q^A=\{\xi,\eta,\psi\}\@$, $\@p\!\@_A=\De L/de{\dot q^A}\@$ and introducing the function
\begin{equation}\label{2.16}
\Routh\/(q^\a\!,q^A\!,{\dot q}^\a\!,p\!\@_A)\,:=\,L\@-\,p\!\@\?_B\,\dot q^B,
\end{equation}
henceforth called the \emph{Routhian\/}, the Lagrange equations take then the form
\begin{subequations}\label{2.17}
\begin{align}
&\d/dt\,\de L/de{\dot q^A}\,-\,\cancel{\de L/de{q^A}}\,=\,0\quad\Longrightarrow\quad p\!\@_A\@=\@\cost              \label{2.17a}\\[4pt]
&\d/dt\,\de\Routh/de{\dot q\@^\a}\,-\,\de\Routh/de{q^\a}\,=\,0                                                      \label{2.17b}
\end{align}
\end{subequations}

In this way, for any assignment of the (constant) values of the conserved momenta $\@p\!\@_A\@$, eqs.~(\ref{2.17b}) are formally identical to a system of three ordinary
Lagrange equations for the unknowns $\@\z\/(t),\th\/(t),\vphi\/(t)\@$.

The implementation of the algorithm is entirely
\linebreak
straightforward: for later convenience, we summarize it into the following
\begin{proposition}\label{Pro2.1}
Given a Lagrangian $\@L=\frac12\@a_{ij}\@\dot q^i\@\dot q^j+\@U$, denote by $\@\{\?q^\a,\;\a=1\And r\?\}\@$ the non cyclic variables, by $\@\{\?q^A,\;A=r+1\And n\?\}\@$
the cyclic ones and by $\@\hat a^{AB}\@$ the inverse of the principal minor $\@a_{AB}\@$ of the matrix $\@a_{ij}\@$ (not to be confused with the minor $\@a^{AB}$ of the
matrix $\@a^{ij}\@$, inverse of $\@a_{ij}\@$). Then, eq.~\eqref{2.16} reads
\begin{equation*}
\Routh\!\@=\!\@\tfrac12\@a_{\a\b}\@\dot q^\a\dot q^\b\!-\tfrac12\@\hat a^{AB}\!\@\big(p_A-a_{A\a}\@\dot q^\a\big)\big(p_B-a_{B\b}\@\dot q^\b\big)+U
\end{equation*}
\end{proposition}
\begin{Proof}
The conclusion follows at once from the equations
\begin{align*}
&p_{\!\@A}\@=\@\de L/de{q^A}\,=\,a_{A\a}\,\dot q\@^\a+\@a_{AB}\,\dot q\@^B                                                  \\
&\Routh\@=L\@-\,\de L/de{q^A}\,\dot q^A\@=\@a_{\a\b}\,\dot q\@^\a\@\dot q\@^\b-\,\@a_{AB}\,\dot q\@^A\@\dot q\@^B+\,U
\end{align*}
The details are left to the reader.
\end{Proof}

In addition to obvious computational advantages, the Routh procedure has also interesting theoretical implications: for example, it helps refining the classification of
the equilibrium configurations, assigning a precise geometrical meaning to the concept of \emph{pseudo--stability}. The idea is formalized by the following
\begin{definition}\label{Def2.1}
Let $\@\B\@$ be a scleronomous system, $\@\hat\A\@$ its velocity space and $\@L\in\mathcal F\/(\hat A)\@$ a Lagrangian.
\\
As in Proposition \ref{Pro2.1}, regard
$\@q^\a,\,\a=1\And r\@$ as non cyclic variables, and $\@q^A,\,A=r+1\And n\@$ as cyclic ones.
\\
Denote by $\@p_{\!\@A}\@$ the kinetic momenta $\@\de L/de{\dot q^A}\@$, and by $\@\Surf_0\subset\hat A\@$ the submanifold described by the equation $p\!\@_A=0$.

An equilibrium configuration $\@q^*=(q^{*1},\ldots,q^{*n})\@$ is then called \emph{pseudo--stable\/} if and only if, for any neighborhood $\@\Eps\@$ of the point
$\@\tilde{q\@}^*=(q^{*1},\ldots,q^{*n},0,\ldots,0)\@$ in $\@\hat\A\@$ there exists a neighborhood $\@\Delta\ni\tilde{q\@}^*\,$ such that, chosen arbitrary initial data
$\@\big(q^i\/(t_0),\dot q^i\/(t_0)\big)\in\Delta\cap\Surf_0\@$, the subsequent evolution of $\@\B\@$ is contained in $\@\Eps\,$.
\end{definition}

Definition \ref{Def2.1} is clearly equivalent to the request that $\@q^*\@$ be a stable equilibrium configuration for the reduced problem based on the Routhian
\eqref{2.16}, restricted to the hypersurface $\@\Surf_0\@$, namely on the function
\begin{multline}\label{2.18}
\Routh_{|\Surf_0}\,=\,\tfrac12\@\big(a_{\a\b}\@-\@\hat a^{AB}\?a_{A\a}\,\@a_{B\b}\big)\@\dot q\@^\a\@\dot q^\b\@+\,U\,:=\\=\,
\tfrac12\,m_{\a\b}\,\dot q^\a\@\dot q^\b\@+\,U\quad
\end{multline}
There exists, therefore, a pseudo--stability criterion, formally identical to the Dirichlet one.

Coming back to the study of the floating rigid body, let us now concentrate on the equilibrium configuration $\@q^*\@$:
\mbox{$\@\xi^*=\eta^*=0\@,\,\z^*=d\@,\,\psi^*=\th^*=\vphi^*=0$}
\footnote%
{By properly choosing the origin $\@\W\@$ and the axes $\@\k_1,\@\k_2\@$, every equilibrium configuration can always be represented in the stated form. This reflects
once again the invariance of the algorithm under the subgroup $\@\E_2\subset\E_3\@$ of rigid motions preserving the plane $\@\z=0\@$.}.

Taking eqs.~(\ref{2.8}\,a,\,b,\,c) and the relation $\@m=\rho\@\?V^*$ into account, it is readily seen that the potential $\@U=m\@g\,\z\@+\@\UB\@$ is indeed
\emph{stationary\/} at $\@q=q^*$.

A sufficient condition for the pseudo--stability of $\@q^*\@$ is therefore the negative--definiteness of the Hessian
\begin{equation}\label{2.19}
\left[\@\begin{matrix}
\plus{14}0\@\SDs U/de\z/de\z \;&\; \SDs U/de\z/de\th \;&\; \SDs U/de\z/de\vphi                               \\[8pt]
\SDs U/de\th/de\z \;&\; \SDs U/de\th/de\th \;&\; \SDs U/de\th/de\vphi                            \\[8pt]
\SDs U/de\vphi/de\z \;&\; \SDs U/de\vphi/de\th \;&\; \SDs U/de\vphi/de\vphi
\plus07\@\end{matrix}\right]\hskip-2pt{\plus0{27}}_{q^*}\!=\,
\left[\@\begin{matrix}
\plus{14}0\@\de\plus03\@Q_{\!\@\?\z}/de\z \,&\, \de\plus03\@Q_{\!\@\?\z}/de\th \,&\, \de\plus03\@Q_{\!\@\?\z}/de\vphi                      \\[7pt]
\@\de\plus03\@Q_{\!\@\?\th}/de\z \,&\,\de\plus03\@Q_{\!\@\?\th}/de\th \,&\, \de\plus03\@Q_{\!\@\?\th}/de\vphi                              \\[7pt]
\plus07\@\de\plus03\@Q_{\vphi}/de\z \,&\, \de\plus03\@Q_{\vphi}/de\th \,&\, \de\plus03\@Q_{\vphi}/de\vphi
\end{matrix}\right]\hskip-2pt{\plus0{27}}_{q^*}\!
\end{equation}

To evaluate the latter, we refer to eq.~\eqref{2.7} and observe that, by an argument identical to the one employed in the proof of Theorem \ref{Teo2.1}, the transport
equation entails the relation
\begin{multline}\label{2.20}
\de Q_k/de{q^r}\,=\,-\rho\@g\@\biggl\{\,\SD R_{3i}/de{q^k}/de{q^r}\@\int_D\@x_i\,\dv\@+\,                                       \\[6pt]
\hskip-7pt+ \int_A\@\biggl(\de\z/de{q^k}\@+\@\de R_{3i}/de{q^k}\,\@x_i\biggr)\@\de P/de{q^r}\cdot\k_3\,d\?S\,\biggr\}
\end{multline}
On the other hand, by eq.~\eqref{2.5} we have
\begin{align*}
\de P/de{q^r}\cdot\k_3\@&=\,\de\vett v_P/de{\dot q^r}\cdot\k_3\@=                                                               \\
&=\,\de/de{\dot q^r}\,\biggl(\dot\z\@+\@\d\/R_{3j}/dt\;x_j\biggr)\,=\,
\de\z/de{q^r}\@+\@\de R_{3j}/de{q^r}\;x_j
\end{align*}
whence, substituting into eq.~\eqref{2.20}
\begin{multline}\label{2.21}
\hskip-2pt\de Q_k/de{q^r}=-\rho\@g\@\biggl[\int_D\@\SD R_{3i}/de{q^k}/de{q^r}\,x_i\,\dv\,+                                      \\[5pt]
+\int_A\biggl(\de\z/de{q^k}\@+\@\de R_{3i}/de{q^k}\,x_i\biggr)\biggl(\de\z/de{q^r}+\de R_{3j}/de{q^r}\,x_j\biggr)\,d\?S\biggr]
\end{multline}

The rest is now straightforward. To state the result in compact form, in addition to the static waterplane $\@A^*$ and to the projection $\@\bar G\@$ of the mass center
$\@G\@$ on $\@A^*\@$, we introduce the following attributes:

\vskip5pt\noindent\tondo
the area $\@\AWP\@$ of $\@A^*$;

\vskip5pt\noindent\tondo
the \emph{floating center\/}  $\@C\@$, defined by the equation
\begin{equation}\label{2.22}
(C-\bar G)\,=\,\frac1\AWP\,\int_{A^*}\@(P-\bar G)\,d\/S \qquad
\end{equation}

\noindent\tondo
the symmetric tensor
\begin{equation}\label{2.23}
S\,=\,\int_{A^*}(P-\bar G)\otimes(P-\bar G)\,d\?S
\end{equation}
expressing a sort of ``second moment'' of the region $\@A^*\@$ with respect to $\@\bar G\@$.

\begin{subequations}\label{2.24}
In body--fixed coordinates, setting $\@S=S_{ij}\,\e_j\otimes\e_j\@$\vspace{1pt}, $(C-\bar G)=\xC\,\e_1+\@\yC\,\e_2\@$, we have the explicit expressions
\begin{align}
& \xC=\frac1\AWP\int_{A^*}x_1\,d\/S\@,\quad \yC=\frac1\AWP\int_{A^*}x_2\,d\/S        \\[5pt]
& S_{ij}=\int_{A^*}x_i\,x_j\,d\/S\@,\,i,j=1,2\,;\;\;\, S_{i3}=S_{3i}=0
\end{align}

In particular, when the plane $\@x_2=0\@$ is a symmetry plane for the body $\@\B\@$, in addition to the vanishing of the components $\@I_{21}\@,I_{23}\@$ of the inertia
tensor we have the obvious simplifications $\@\yC=0\@$, $S_{12}=0\@$.
\end{subequations}

After these preliminaries, let us now complete the evaluation of the right--hand side of eq.~\eqref{2.19}.
\\
To this end we observe that, in the configuration $\@q^*$, eq.~\eqref{2.1} entails the relations \vspace{3pt}
\begin{align*}
&\left.\de\/R_{3i}/de\z\right|_{q^*}\! x_i=0\@,\,\; \left.\de\/R_{3i}/de\th\right|_{q^*}\!x_i=-\@x_1\@,\,\;
\left.\de\/R_{3i}/de\vphi\right|_{q^*} x_i=x_2                                                                                                      \\[8pt]
&\left.\SD\/\/R_{3i}/de /de\th\right|_{q^*}\! x_i=\left.\SD\/\/R_{3i}/de /de\vphi\right|_{q^*}\! x_i\@=\@-\@x_3\@,\,\;
\left.\SD\/R_{3i}/de\th/de\vphi\right|_{q^*}\!x_i\@=\@0                         \\[8pt]
& \left.\SD\/R_{3i}/de\z/de{q^k}\right|_{q^*} x_i\@=\@0
\end{align*}

From these, denoting by $\@\H\@$ the Hessian \eqref{2.19} and\vspace{1pt} by $\@z^*_B=(B^*-G)\cdot\e_3=\frac1{\plus60 V^*}\@\int_D\@x_3\,\dv\@$\vspace{1pt} the third
coordinate of the buoyancy center at equilibrium, we get the expression
\begin{equation}\label{2.25}
\H\,=\,\rho\@g
\left[\@\begin{matrix}
\plus{14}0\@-\AWP & \AWP\,\xC & 0                      \\[9pt]
\!\AWP\,\xC \,&\,V^*z^*_B-S_{11} & 0                              \\[9pt]
\@0 & 0 & \!\!V^*z^*_B-S_{22}\,
\plus07\end{matrix}\right]
\end{equation}

\medskip
In particular, due to the positivity of $\@\rho,\@g,\?\AWP\@$, the negative--definiteness of the matrix \eqref{2.25} is equivalent to the pair of conditions
\begin{equation}\label{2.26}
S_{22}\@-\@V^*z^*_B\@>\@0\,,\qquad S_{11}\@-\@V^*z^*_B\@>\@\AWP\,\xC^2
\end{equation}

Let us finally recall that, in marine engineering, it is customary to introduce the transverse and longitudinal metacentric heights, respectively defined as
\begin{equation}\label{2.27}
\GMT\@:=\@\frac{S_{22}}{V^*}\@-z^*_B\,,\qquad\;\, \GML\@:=\@\frac{S_{11}}{V^*}\@-z^*_B
\end{equation}
with the understanding that, if the origin of the body--fixed frame is not located at the mass center $\@G\@$, the quantity $\@z^*_B\@$ is replaced by the projection
$\@(B^*-G)\cdot\e_3\@$.

With these definitions, denoting by $\Delta:= m\@g=\rho\@g\@V^*$ the displacement of $\@\B\@$, eq.~\eqref{2.25} reads
\begin{equation}\label{2.28}
\H\,=\,
\left[\@\begin{matrix}
\plus{14}0 -\@\rho\@g\AWP &\;\rho\@g\AWP\,\xC & 0                                                     \\[9pt]
\rho\@g\AWP\,\xC \,&\;-\@\Delta\,\GML & 0                                                           \\[9pt]
\@0 & 0 & -\@\Delta\,\GMT\,
\plus07\end{matrix}\right]
\end{equation}
while the conditions \eqref{2.26} for the pseudo--stability of the configuration $\@q^*\@$ acquire the standard form \cite{Lewandowski}
\begin{equation}\label{2.29}
\GMT\@>\@0\,,\qquad\;\,\Delta\,\GML\@>\@\rho\@g\@\AWP\,\@\xC^2
\end{equation}

\section{Small oscillations about the equilibrium configuration}\label{S4}
Consistently with Definition \ref{Def2.1}, the \emph{small oscillations\/} of a scleronomous system $\@\B\@$ around a pseudo--stable equilibrium configuration $\@q^*\@$
are defined as evolutions in which all conserved kinetic momenta $\@p_A\@$ are zero and all deviations $\@\eta^\a\/(t)=q^\a\/(t)-q^{*\@\a}$, $\;\dot\eta^\a\/(t)=\dot
q^\a\/(t)\@$ are small, thus justifying the replacement of the exact equations of motion with corresponding \emph{linearized\/} ones.

The algorithm is entirely standard: the function \eqref{2.18} is developed up to second order in the deviations, yielding a corresponding \emph{approximate Routhian\/}
\begin{equation*}
\tilde\Routh\/(\eta^\a,\dot\eta^\a)\,=\,\tfrac12\,m_{\a\b}\,\dot\eta^\a\@\dot\eta^b\?-\,\tfrac12\,c_{\a\b}\,\eta^\a\@\eta^b
\end{equation*}
with $\@m_{\a\b}=m_{\a\b}(q^*)\@$ and $\@c_{\a\b}=-\@\H_{\a\b}\@$ identical to the opposite of the Hessian \eqref{2.25}.
The linearized equations of motion are then the Lagrange equations
\begin{equation}\label{2.30}
\d/dt\,\de\tilde\Routh/de{\dot q\@^\a}\,-\,\de\tilde\Routh/de{q^\a}\,=\,0
\end{equation}

Due to the positive definiteness of both matrices $\@c_{\a\b}\@$~and~$\@m_{\a\b}\@$, the solutions of eqs.~\eqref{2.30} are linear combinations of harmonic oscillations,
with frequencies
\linebreak
$\nu=\sqrt{\l}/2\@\pi\@$ determined by the eigenvalue equation
\begin{equation}\label{2.31}
\det (c_{\a\b}\@-\@\l\,m_{\a\b})\,=\,0
\end{equation}

Starting from eq.~\eqref{2.13} and applying the prescriptions outlined in Proposition~\ref{Pro2.1}, a straightforward calculation yields the evaluation
\begin{equation*}
m_{\a\b}\,=\,\left[\@\begin{matrix}
\plus{10}0\,m & \quad 0 & 0         \\[4pt]
0 & \quad I_{22} & 0  \\
 0  & \quad 0 &\; \Frac{\plus04 I_{11}^{\phantom 2}\@I_{33}^{\phantom 2}\,-\,I_{13}^{\@2}}{\plus80I_{33}}\,
\plus0{10}\end{matrix}\right]
\end{equation*}

\smallskip\noindent
Eq.~\eqref{2.31} takes therefore the form
\begin{equation*}
\det\!\left[\begin{matrix}
\plus{10}0\text{\footnotesize{$\rho\@g\AWP-\l\@m$}} & \@\text{\footnotesize{$-\rho\@g\AWP\,\xC$}} &  \text{\footnotesize{$0$}}                            \\[5pt]
\text{\footnotesize{$\!\!-\rho\@g\AWP\@\xC$}} &\;\text{\footnotesize{$\Delta\@\GML-\l\@I_{22}$}}\, & \text{\footnotesize{$0$}}                            \\[2pt]
\@\text{\footnotesize{$0$}} &\text{\footnotesize{$0$}} &\hskip-12pt\text{\footnotesize{$\Delta\@\GMT-\l$}}\,\frac{I_{11}\@I_{33}\,-\,I_{13}^{\@2}}{I_{33}}
\plus0{10}\end{matrix}\right]\!=0
\end{equation*}
relating the frequencies of the normal modes of vibration to the geometric and material properties of the body.


\end{document}